\documentclass[a4paper,fleqn,usenatbib]{mnras}

\usepackage{graphicx}	
\usepackage{amsmath}	
\usepackage{amssymb}	
\usepackage{rotating}
\usepackage{pdflscape}
\usepackage{longtable}
\usepackage{afterpage}
\usepackage[T1]{fontenc}

\title[ANN classification of M1:2 asteroids]
    {Artificial Neural Network classification of asteroids in the M1:2 mean-motion resonance with Mars}

\author[V. Carruba, S. Aljbaae, R. C. Domingos, W. Barletta]{V. Carruba$^{1}$\thanks{E-mail: valerio.carruba@unesp.br}, S. Aljbaae$^{2}$, R. C. Domingos$^{3}$, W. Barletta$^{1}$\\
  $^{1}$S\~{a}o Paulo State University (UNESP), School of Natural Sciences and Engineering, Guaratinguet\'{a}, SP, 12516-410, Brazil \\
  $^{2}$National Space Research Institute (INPE), Division of Space Mechanics and Control, C.P. 515, 12227-310, S\~{a}o Jos\'e dos Campos, SP, Brazil \\
  $^{3}$S\~{a}o Paulo State University (UNESP), S\~{a}o Jo\~{a}o da Boa Vista, SP, 13876-750, Brazil 
}

\date{Accepted 2021 March 25. Received 2021 March 25; in original form 2021 March 2}
\pubyear{2021}

\begin{document}
\label{firstpage}
\pagerange{\pageref{firstpage}--\pageref{lastpage}} 
\maketitle

\begin{abstract}

  Artificial neural networks ({\it ANN}) have been successfully used in the
  last years to identify patterns in astronomical images.  The use
  of {\it ANN} in the field of asteroid dynamics has been, however, so far
  somewhat limited.  In this work we used for the first time {\it ANN} for
  the purpose of automatically identifying the behaviour of asteroid orbits
  affected by the M1:2 mean-motion resonance with Mars.  Our model was able
  to perform well above 85\% levels for identifying images of asteroid
  resonant arguments in term of standard metrics like {\it accuracy,
    precision} and {\it recall}, allowing to identify the orbital type of
  all numbered asteroids in the region. Using supervised machine learning
  methods, optimized through the use of genetic algorithms, we also predicted
  the orbital status of all multi-opposition asteroids in the area.  We
  confirm that the M1:2 resonance mainly affects the orbits of the Massalia,
  Nysa, and Vesta asteroid families.
  
\end{abstract}


\section{Introduction}
\label{sec: intro}

During the last five years, machine learning and deep learning have been
more and more been used in the field of asteroid dynamics.  Among the
latest application, supervised methods of machine learning
have been used to identify the population of
asteroids in three-body mean-motion resonances \citep{2017MNRAS.469.2024S},
new members of known asteroid families \citep{Carruba_2020}, and
asteroids groups inside the $z_1$ and $z_2$ secular resonances
\citep{Carruba_2021}, among others.  Deep learning in the form
of artificial neural networks has been recently used for identifying
members of asteroid families \citep{Vujicic_2020}.
While several applications of artificial neural networks exist
in other astronomy fields for the purpose of identifying images, like, for
instance, methods to identify different types of galaxies clusters
\citep{Su_2020}, to our knowledge such methods have not yet been applied for
asteroid dynamics problems.

Here we attempt for the first time to use artificial neural networks for
automatically identifying the behaviour of asteroids near the two-body
M1:2 mean-motion resonance with Mars.  As discussed by other authors
\citep{Gallardo_2011}, three types of
orbits are possible near resonance: {\it libration}, where the resonant
argument of the resonance, which we will define in
section~(\ref{sec: m12_dyn}), oscillates around an equilibrium point,
{\it circulation}, where the resonant argument cover the whole range
of values from $0^{\circ}$ to $360^{\circ}$, and {\it switching} orbits,
where the resonant argument alternates phases of libration and circulation.
In previous works, the classification of the type of orbits on which an
asteroid resides was either performed manually, by visually inspecting
  the time behaviour of the resonance argument (see, for instance,
  \citet{2018P&SS..157...72C} and references therein), or by using
  automatic algorithms for the same purpose
  \citep{2018Icar..304...24S,2014Icar..231..273G,2016Icar..274...83G}.
In this work, we use artificial
neural networks for classifying an asteroid's orbital type
for all numbered asteroids in the region affected by this resonance.

We then employed genetic algorithms to select the best performing machine
learning supervised method, to predict the labels of
multi-opposition objects in the area.  Multi-opposition asteroids are
asteroids that have been observed at several oppositions to the Sun from
Earth, whose orbits are somewhat well-established.  Once the orbit is confirmed,
an asteroid receives an identification number and becomes a numbered asteroid,
like 2 Vesta, 4  Pallas, 10 Hygiea, among others.  Since the orbits
of multi-oppositions asteroids are not as well-established as those
of numbered bodies, here we used the labels of numbered asteroids
to predict those of the multi-oppositions objects. Finally, we verified which
local asteroid families are most affected by this dynamical resonance,
to see if our results are consistent with those in the literature.
We start our analysis by revising the dynamical properties of
asteroids in the region.

\section{The population of asteroids in the M1:2 resonance: dynamics}
\label{sec: m12_dyn}

The population of asteroids inside the M1:2 mean motion resonance with Mars
has been the subject of a study by \citet{Gallardo_2011}, that investigated
the dynamical, physical and evolutionary properties of these asteroids.
Here we will briefly summarize the dynamical characteristics of this
population, and distinguish between the types of orbits possible in the
orbital region affected by this resonance.
Figure~(\ref{Fig: m12_prop_ae}) displays an $(a,e)$ projection of 9457
numbered asteroids in the range in $a$ from 2.411 to 2.426 au.  Values
of synthetic proper elements for asteroids in the region were obtained from
the {\it Asteroid Families Portal} ($AFP$, \citet{Radovic_2017}, accessed on
August 1st, 2020). Synthetic proper elements are constant of the motion on
timescales of millions of years and are obtained as the outcome of numerical
simulations, using methods described in \citet{Knezevic_2003}. The V-shaped
region at the resonance centre is associated with the M1:2 resonance. The
higher number concentration of objects at the edge of the V-shape is caused
by the phenomenon called ``resonance stickiness'' \citep{Malishkin_1999}.
\citet{Gallardo_2011} define two main resonant arguments for this resonance.
$\sigma$ is given by:

\begin{figure}
  \centering
  \centering \includegraphics[width=3.0in]{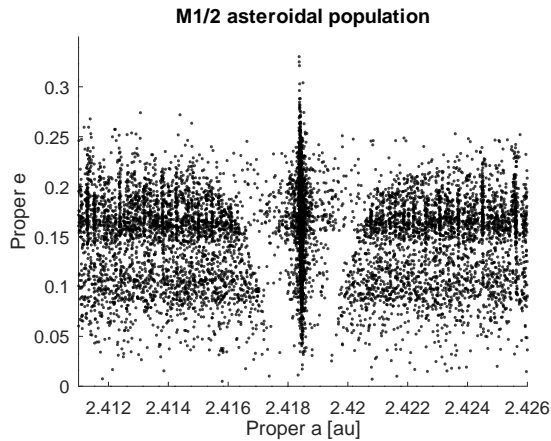}
  \caption{Proper $(a,e)$ distribution for asteroids in the orbital region
  of the M1:2 mean-motion resonance.}
\label{Fig: m12_prop_ae}
\end{figure}

\begin{equation}
\sigma=2 \lambda -{\lambda}_{M}-\varpi,
\label{eq: sigma}
\end{equation}

\noindent where $\lambda = M+\Omega+\omega$ is the mean longitude,
$\varpi=\Omega+\omega$, with $\Omega$ the longitude of the node, $\omega$
the argument of pericentre, and where the suffix $M$ identifies the planet
Mars. ${\sigma}_1$ is defined as:

\begin{equation}
{\sigma}_1=2 \lambda -{\lambda}_{M}-{\varpi}_{M}.
\label{eq: sigma_1}
\end{equation}

\begin{figure*}
\centering
\includegraphics[width=3.5in]{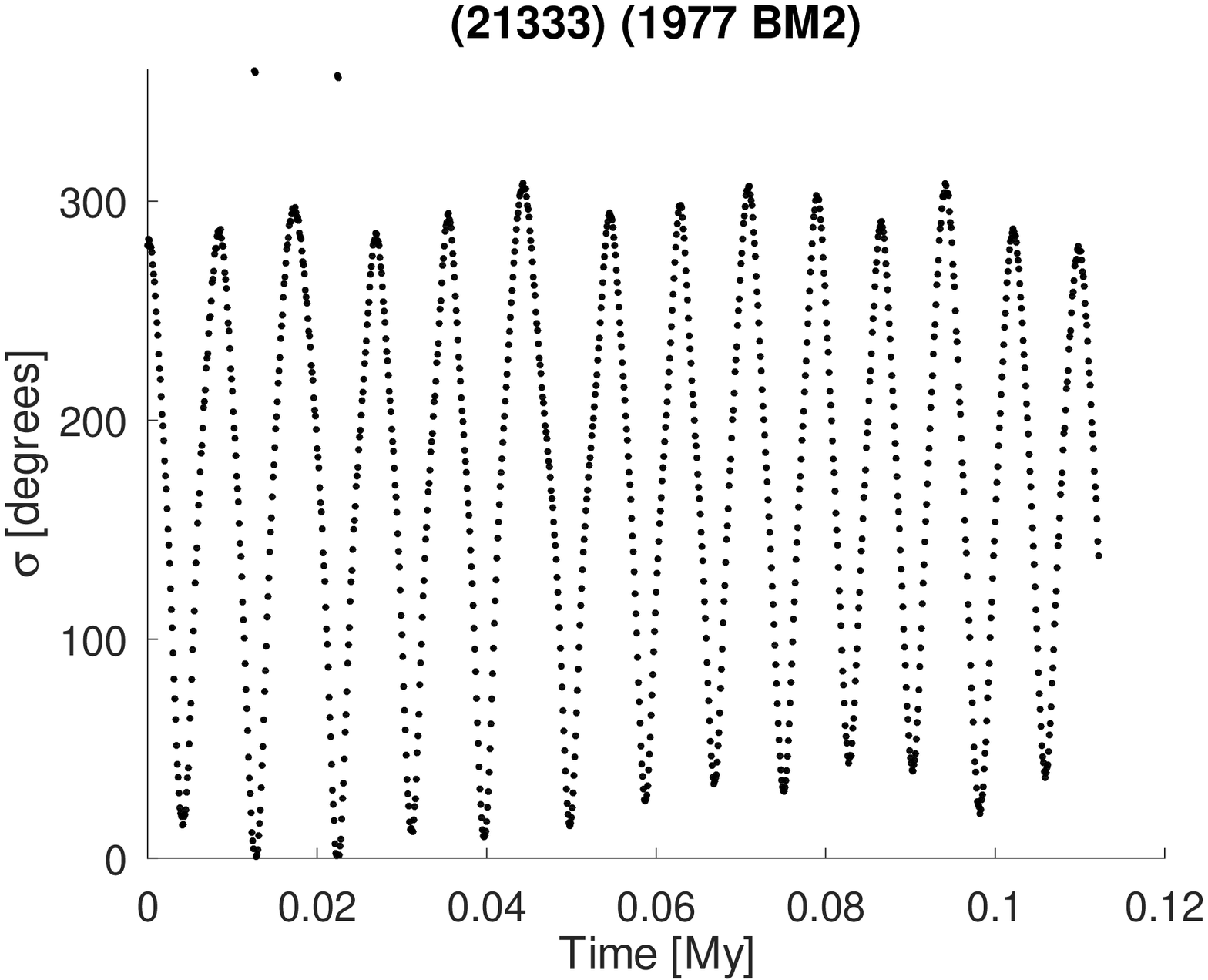}
\includegraphics[width=3.5in]{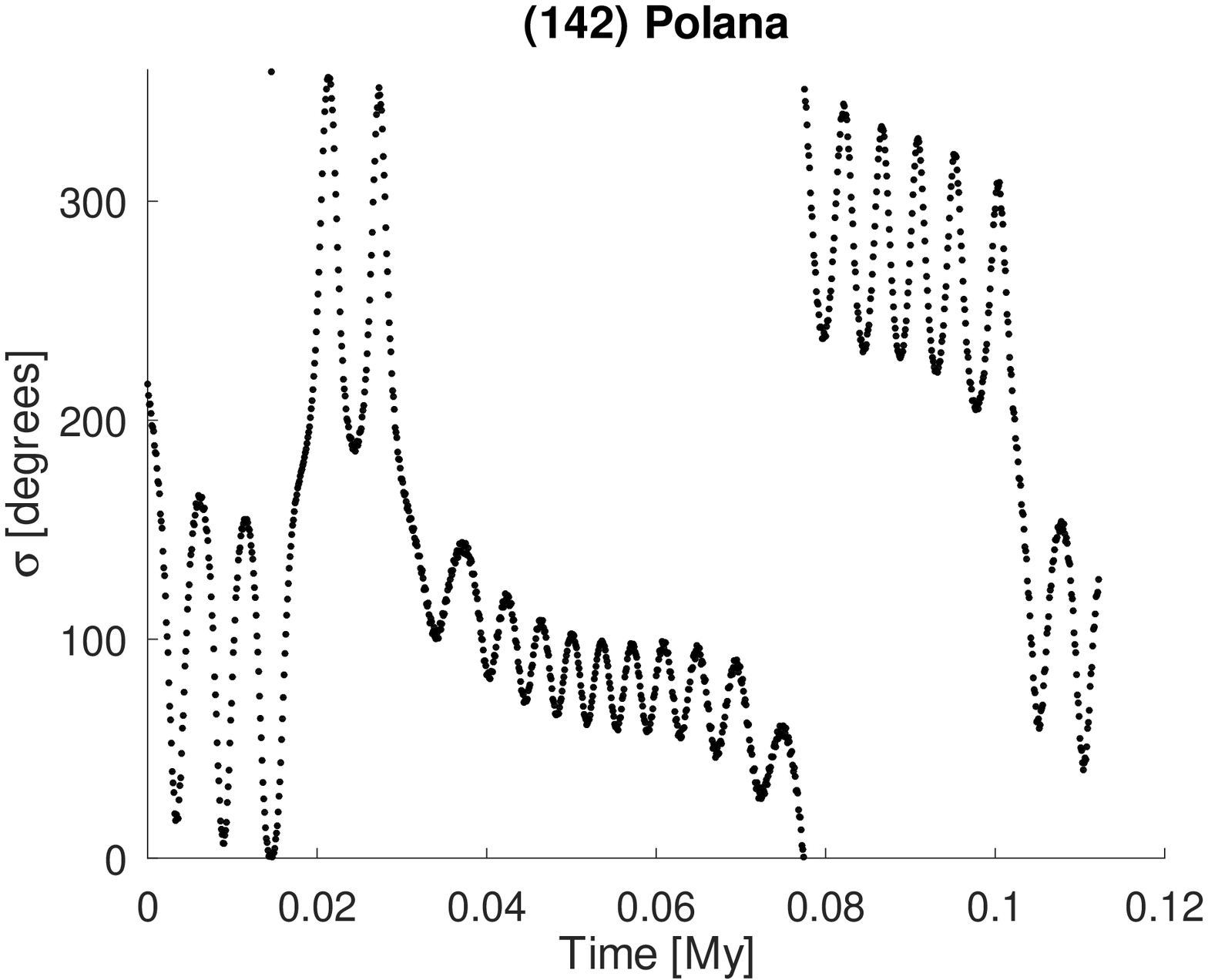}
\includegraphics[width=3.5in]{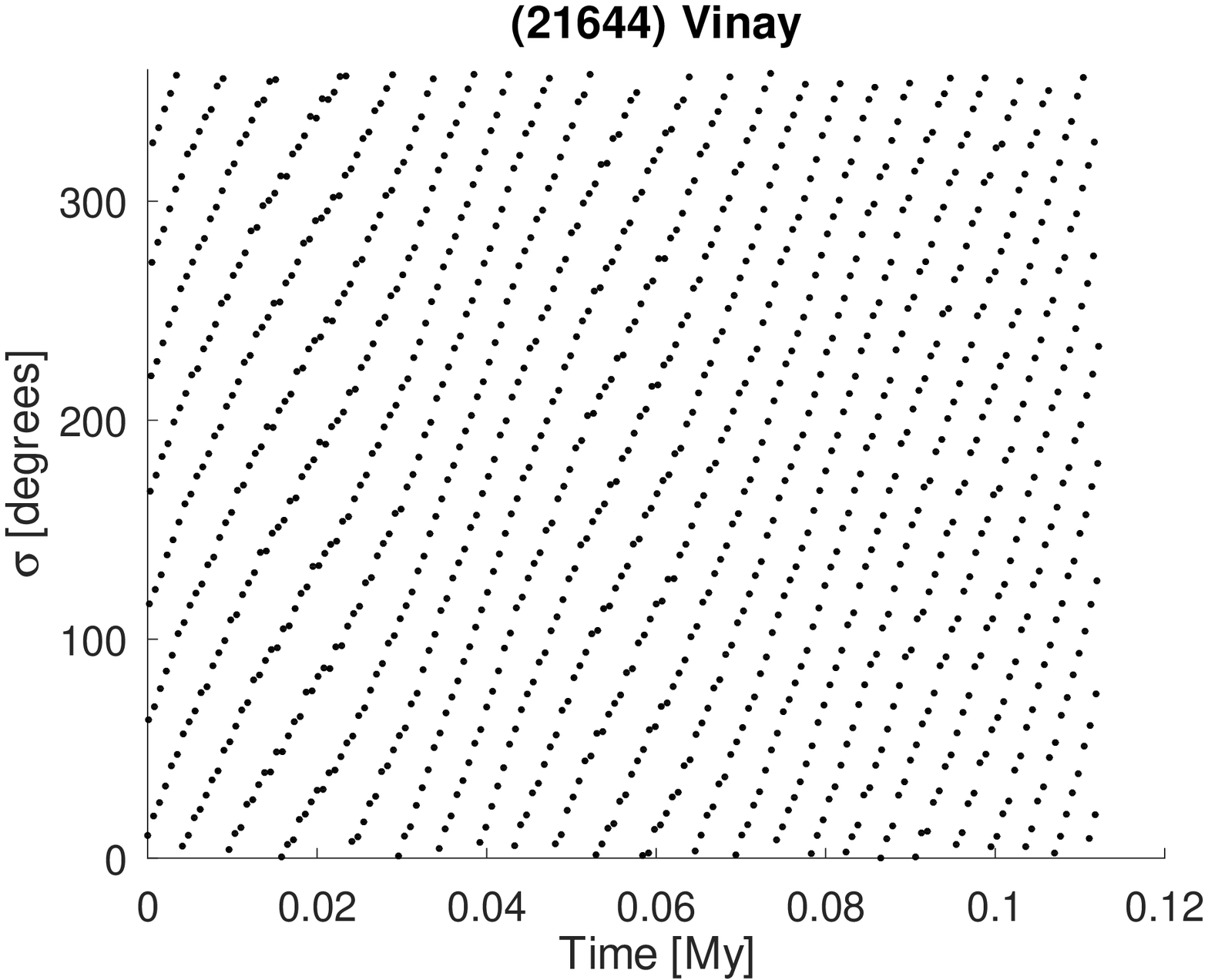}
\caption{The resonant angles $\sigma$, as defined by equation~(\ref{eq: sigma})
  as a function of time for asteroids on librating, switching and
  circulating orbits of the M1:2 resonance.}
\label{Fig: polana_res_ang}
\end{figure*}

The orbital behaviour of asteroids in the affected region can be identified
by studying the time dependence of these two angles.  As previously discussed,
asteroids for which the critical arguments cover the whole range of values,
from $0^{\circ}$ to $360^{\circ}$, are on {\it circulating} orbits.  If the
argument oscillates around an equilibrium point we have a {\it librating}
orbit. Whether the argument alternates phases of libration and
circulations, or
switch between different equilibrium points, we have a {\it switching} orbit,
as defined in this work.  We identify the orbital types of asteroids by
performing a 100000 yr simulation with the Burlisch-Stoer integrator of the
{\it SWIFT} package \citep{levison_1994}. We use a time step of 1 day, a
tolerance ($EPS$) equal to $10^{-8}$, and integrated the asteroids under the
influence of all planets. None of the asteroids in our sample is a
Mars-crosser or susceptible to experience close encounters with planets,
which justifies the use of a Burlisch-Stoer integrator for this study.
Figure~(\ref{Fig: polana_res_ang}) show the resonant
argument for three asteroids in each of the three classes.  As discussed by
\citet{Gallardo_2011}, since the M1:2 is an external resonance, unusual
equilibrium points for the $\sigma$ argument, like one at $100^{\circ}$, can
occur.  The main equilibrium point for the ${\sigma}_1$ argument is around
$0^{\circ}$.

\begin{figure*}
\centering
\includegraphics[width=3.0in]{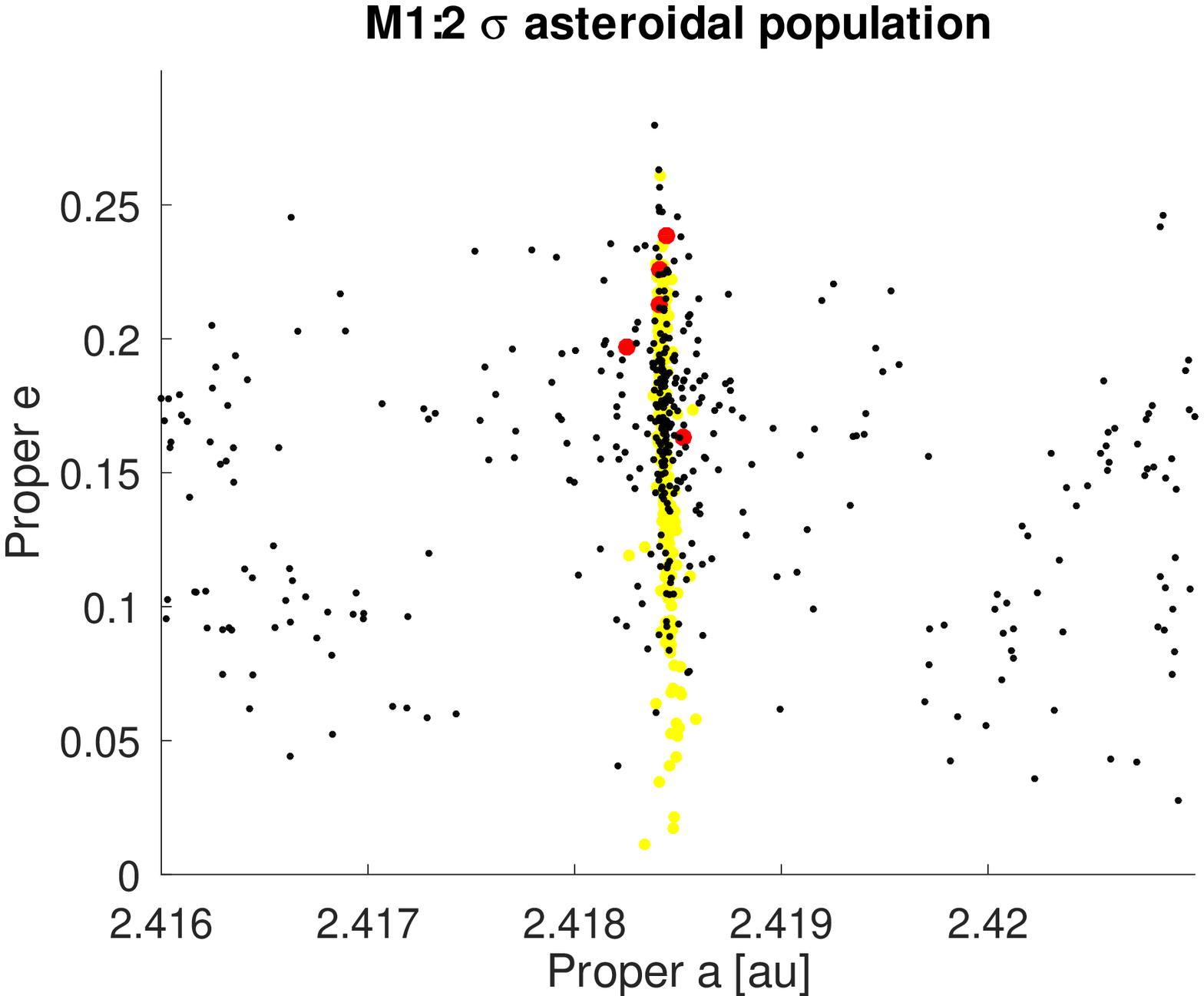}
\includegraphics[width=3.0in]{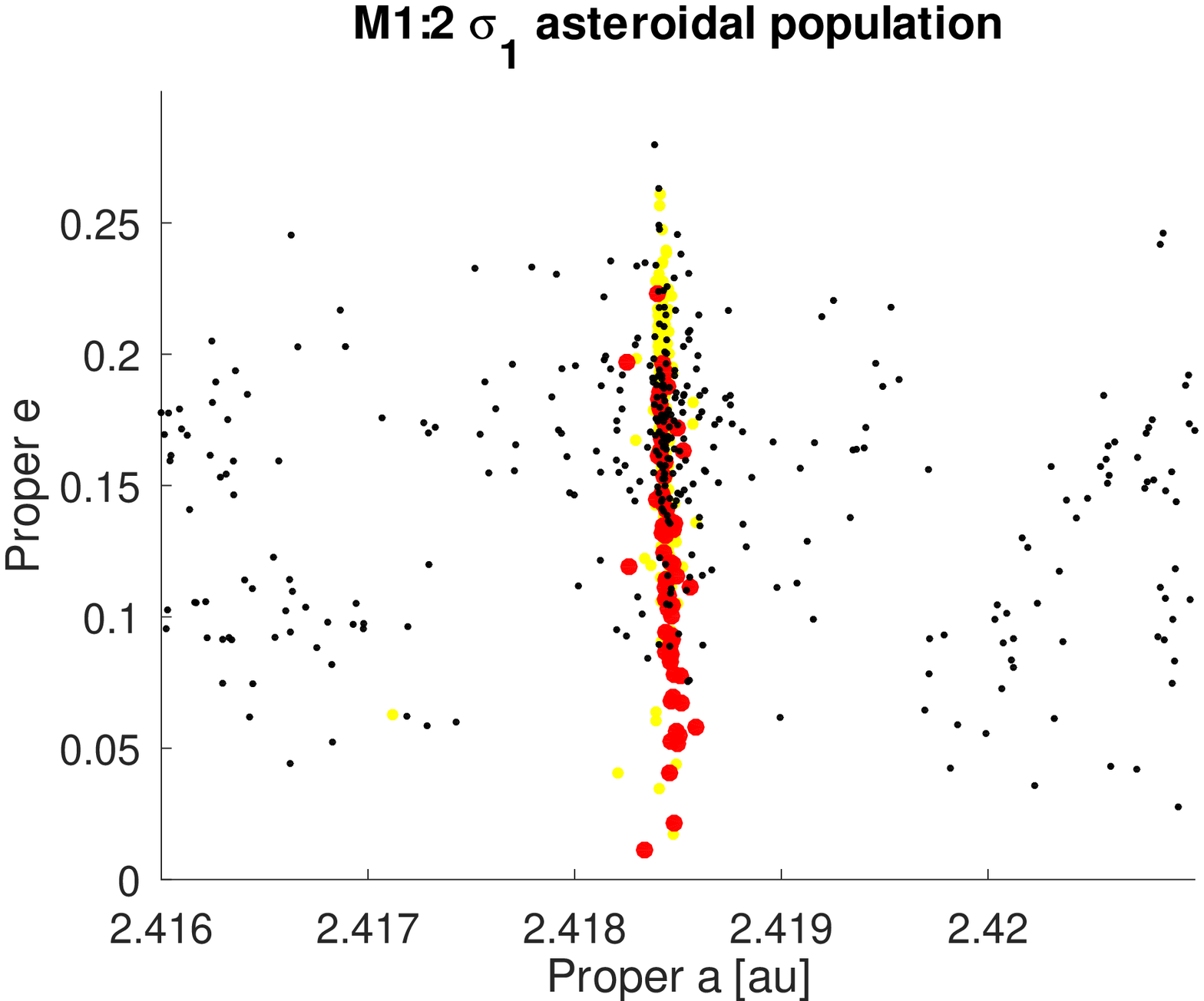}
\caption{A proper $(a,e)$ projection of asteroids in the region of
  the M1:2 resonance.  The left panel shows the orbital behaviour for
  the $\sigma$ resonant argument colour-coded as follows: red full circles
  are librators, yellow full circles are oscillators, and black dots
  are circulators.  The right panel does the same, but for the ${\sigma}_1$
  resonant argument.}
\label{Fig: m12_expl_anal}
\end{figure*}

Using this simulation set-up, we integrated 1000 asteroids in the orbital
region of the M1:2 mean-motion resonance.  Figure~(\ref{Fig: m12_expl_anal})
shows an $(a,e)$ projection of these asteroids, colour-coded for the
behaviour of the $\sigma$ (left panel) and ${\sigma}_1$ resonant argument.
The main difference between the two cases is the fraction of asteroids
in pure librating states.  For the case of $\sigma$, there were just
4 librators (0.4\%) and 202 oscillators (20.2).  For ${\sigma}_1$,
there were 69 librators (6.9\%) and 185 oscillators (18.5\%).  Pure $\sigma$
librators tend to be much rarer than pure ${\sigma}_1$ ones.  Since in this
work we are interested in treating a multi-class problem, rather than
a binary one, from now on we will focus our study on the case
of the ${\sigma}_1$ resonant arguments.

\section{Artificial Neural Networks}
\label{sec: ANN}

At the time that we carried out this study, there were 6440
numbered and multi-opposition asteroids in the region of the M1:2
mean-motion resonance.  Analyzing resonant arguments for each of the
asteroids in the region may be a very tiring and time-consuming endeavour, if
performed manually.  Automatic approaches not based on machine-learning
have been developed in the last years to solve this problem
\citep{2018Icar..304...24S,2014Icar..231..273G,2016Icar..274...83G}.
Here this task will be performed by using artificial
neural networks ({\it ANN}).   The human brain classifies images by
converting the light received by the eye's retina into electrical signals,
that are then processed by a hierarchy of connected neurons to identify
patterns.

\begin{figure}
  \centering
  \centering \includegraphics[width=3.0in]{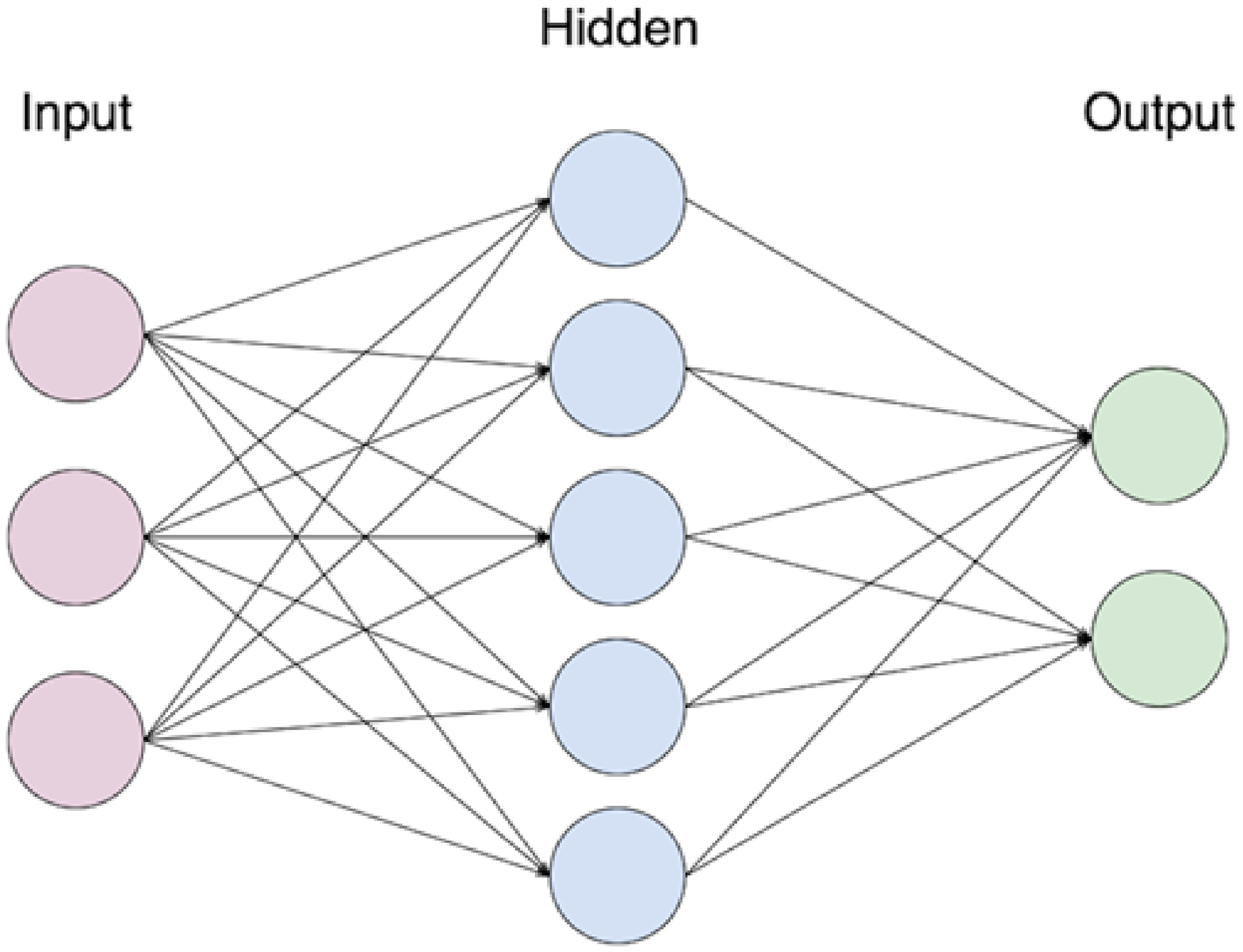}
  \caption{A simple architecture for an {\it Artificial Neural Network}.
    The network has 3 neurons in the input layer, 5 in the hidden one
    and two in the output layer.}
\label{Fig: Simple_ANN}
\end{figure}

Artificial neuron networks mimic the neurons web in a biological brain.
Each artificial neuron can transmit a signal to other neurons.  This signal,
which is usually a real number, can be processed, and the signal coming
out of each neuron is computed as a non-linear function of the inputs.
A basic architecture for {\it ANN} consists of an input and an output layers,
with the possible presence of one or more hidden layers between them to
improve the model precision.  Generally speaking, input layers will look
for simpler patterns, while output layers will search for more complex
relationships. Figure~(\ref{Fig: Simple_ANN}) shows the architecture of a
simple {\it ANN}, with 3 neurons in the input layer, 5 in the hidden stratus,
and 2 in the output layer.  Each neuron will perform a weighted sum, $W_S$,
given by:

\begin{equation}
W_S= \sum_{i=1}^{n} w_i X_i,
\label{eq: W_S}
\end{equation}

\noindent where n is the number of input to process, $X_i$ are the signals
from other neurons, and $w_i$ are the weights.  {\it ANN} will optimize
the values of the weights during the learning process.  On the
weighted sum $W_S$, {\it ANN} will apply an activation function.  For images
classifications, one of the most used activation function is the
``{\it relu}'', defined as:

\begin{equation}
y=max(W_S,0),
\label{eq: relu}    
\end{equation}

\noindent which will produce as an outcome the weighted sum itself $W_S$, if
that is a positive number, or 0, if $W_S$ has a negative value.
As a next step, the loss function must be applied to all the weights
in the network through a back-propagation algorithm.  A loss function
is usually calculated by computing the differences between the predicted
and real output values. An example of a loss function is the mean squared
error, defined as:

\begin{equation}
  C= \frac{1}{2}\sum_{j=1}^{n}(y_j-{\overline{y}}_j)^2,
  \label{eq: loss_MSE}
\end{equation}

\noindent where ${\overline{y}}_j$ is the expected value of the j-th outcome.
For classification problems with multiple classes, with single classes
identified by numbers, like the problem that we
will discuss in this paper, the {\it sparse\_categorical\_crossentropy}
loss function is generally used.  Interested readers can find more information
on the definition and use of this and other loss functions in the {\it Keras}
documentation (https://keras.io/, \citet{Chollet_2018}).
Once the loss function has been computed, the next step is to find its minimum,
to optimize the values of the weights.  Optimization algorithms
find the gradient of the loss function and update the weights in the {\it ANN}
based on this result.  In this work, we will use the {\it Adam} optimizer
\citep{Kingma_2015}.

{\it ANN} use initial values of weights near zero.  The first row of data is
provided as input and processed through the network.  The prediction of the
network is compared to the real result, and the optimization of the cost
function updates the values of the weights.  This procedure is then repeated
for all data, or, in some cases, for a subset, also called batch. An epoch
is completed when the training procedure is finished for all the observations.
This whole process can then be repeated for other epochs, to improve the quality
of the predictions.

Interested readers could find more information about the use of {\it ANN} in
artificial intelligence in \citep{Lecun_2015}, or in the recent work on
the application of {\it ANN} to the identification of asteroids belonging to
asteroid families by \citep{Vujicic_2020}, and references
therein.  In the next subsection we will discuss applications of
{\it ANN} for the classification of images.

\subsection{Applications of {\it ANN} to M1:2 resonant arguments images}
\label{sec: ANN_M12}

Here we used the {\it Keras} implementation
of {\it ANN}, which is also based on the {\it Tensorflow Python} software
package \citep{Chollet_2018}.  The process used in this work is the
following:
\begin{enumerate}
\item The asteroid orbits are integrated under the gravitational influences
  of the planets.
\item We compute the resonant arguments
\item Images of the time dependence of resonant arguments are drawn
\item The {\it ANN} trains on the training label image data
\item Predictions on the test images are obtained, and images of
  the test data, with their classification are produced.
\end{enumerate}

The last step of producing images for the test data, with the proposed
classification, is performed to make a visual confirmation by the user
easier. The theory behind steps (i) and (ii) was discussed in
section (\ref{sec: m12_dyn}).  Here we will focus on steps (iii), (iv) and
(v).
100$\times$ 100 pixel images of resonant
arguments of the M1:2 resonance were stored and
pre-processed before applying our model in step (iii).  Each image pixel values fall in
the range from 0 to 255. Before feeding the images to our model, we
normalized the pixel values to a range between zero and one, to help the ANN
to learn faster. The choice of the image resolution was a compromise between
not exceeding the computer memory available in our machines while still
having a resolution sufficient for the {\it ANN} to successfully work.  

\begin{figure}
  \centering
  \centering \includegraphics[width=3.0in]{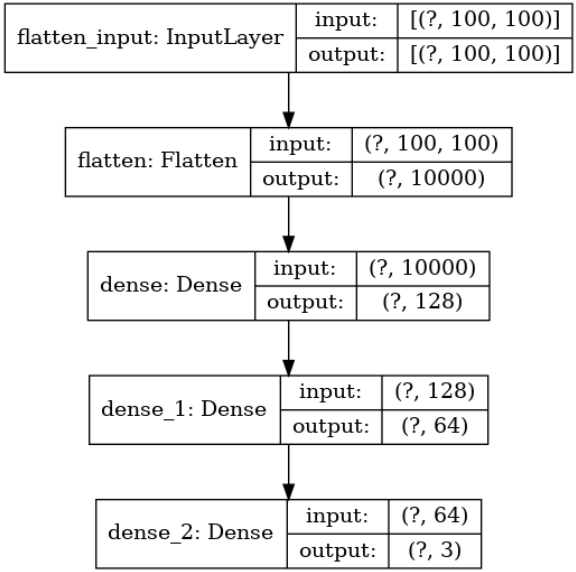}
  \caption{Neural network structure of the model used for classifying resonant
    arguments.  We used a flatten layer, an inner layer, a hidden layer, and
    an output layer for final classification.}
\label{Fig: model_structure}
\end{figure}

To identify resonant argument images, we created
a four-layer model with a flatten, an inner, a hidden, and an output layers.
The architecture of the model is displayed in
figure~(\ref{Fig: model_structure}).  The flatten layer will transform
the image matrices into arrays.  The inner layer will look for
simpler patterns in the arguments images, while the hidden layer,
with half the neurons of the inner one, will
search for more complex features.  The output layer, with three nodes,
will perform the final classification for the three possible classes:
{\it circulation, switching} and {\it librating} orbits.

To quantitatively classify the outcome of {\it ANN}, it is often useful
to compute values of metrics. Some of the most commonly used metrics for
classifications problems are the {\it accuracy}, {\it recall} and
{\it precision}.  For a given class of orbits, we define True-Positive
({\it TP}) as the number of images successfully identified as belonging
to that class by both the observer and the {\it ANN}
model. True-Negative ({\it TN}) are the number of images that both methods
identify as non-belonging to a given class.  False-Positive ({\it FP}) are
images classified as belonging to a class just by the {\it ANN} method.
Finally, False-Negative ({\it FN}) are the images not classified to belong to
a class just by the {\it ANN} approach.  Values of $TP,TN,FP$ and $FN$ can be
obtained by computing the confusion matrix on the images real and predicted
labels.  With these definitions, $accuracy$ \citep{FAWCETT2006861} is given by:

\begin{equation}
  accuracy= \frac{TP+TN}{TP+TN+FP+FN},
\label{eq: accuracy}
\end{equation}

\noindent {\it Recall}, also known as {\it Completeness} in
\citet{Carruba_2020}, is given by:

\begin{equation}
  Completeness= \frac{TP}{TP+FN},
\label{eq: completeness}
\end{equation}

\noindent {\it Precision}, also known as {\it Purity} in \citet{Carruba_2020}, is defined as:

\begin{equation}
  Purity= \frac{TP}{TP+FP}.
\label{eq: purity}
\end{equation}

\noindent While accuracy can yield information on the efficiency of
the algorithm as a whole, {\it Completeness} may inform on the ability
of the method to efficiently retrieve the actual population of a given
class, while {\it Purity} is related to the ability of the model not to
include too many false positives ($FP$).  The optimal model should be trained
to give a trade-off between values of {\it Completeness} and of {\it Purity}.
\citet{Carruba_2020} recently introduced a {\it Merit} metrics that can
automatically perform this trade-off, by giving larger weight to
  Purity.  This new metrics is defined as:

\begin{equation}
  Merit = \frac{1}{\sqrt{5}}\sqrt{{(Completeness)}^2+4\times{(Purity)}^2}.
  \label{eq: Merit}
\end{equation}
     
\noindent In \citet{Carruba_2020}, a higher weight was given to {\it Purity}
with respect to {\it Completeness} because this metric was more relevant
to that work.  Different definitions of {\it Merit} can be made, depending
on the type of problem to be studied.

\begin{figure}
  \centering
  \centering \includegraphics[width=3.0in]{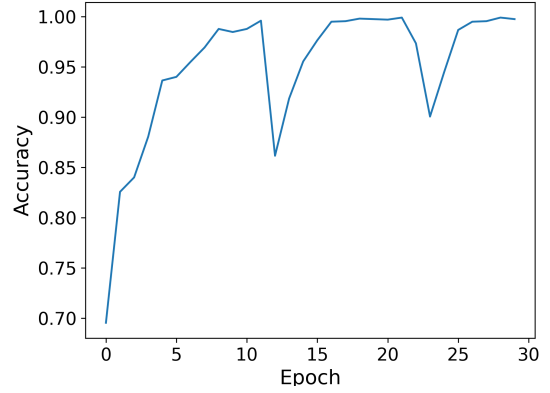}
  \caption{Dependence of accuracy as a function of epoch.}
\label{Fig: accuracy_hist}
\end{figure}

As discussed in section~(\ref{sec: ANN}), the training of {\it ANN}
can be performed for an arbitrary number of times, or {\it epochs}, to
optimize the quality of the predictions.  Figure~(\ref{Fig: accuracy_hist})
displays a plot of {\it accuracy}, as defined by equation~(\ref{eq: accuracy})
as a function of epoch for the training of a neural network with a training
set of 1000 images and a test set of 200 images.  Values of {\it accuracy}
improve as a function of time, but there may be fluctuations from one epoch
to the other, as shown in figure~(\ref{Fig: accuracy_hist}) for epochs 
12 to 13, and 22 to 23.  To avoid using non-optimal weights
for the {\it ANN}, we use a callback instruction, as implemented by
{\it Keras}, during the training to save the weights of each model, and
automatically upload for the model predictions the weights associated with
the best outcome in term of {\it accuracy}.

\begin{figure*}
  \centering
  \centering \includegraphics[width=5.5in]{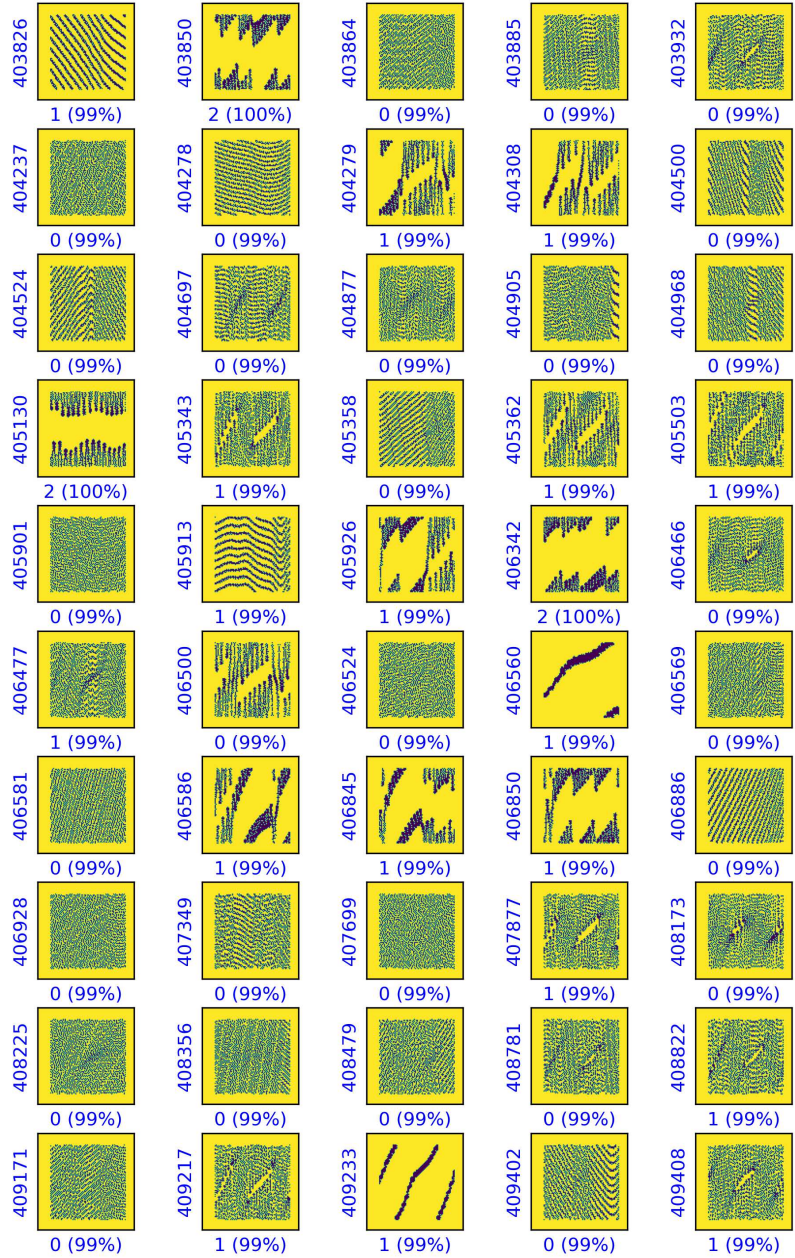}
  \caption{A set of 50 images with the prediction from the {\it ANN} model.
    The percentage values identify the confidence level with which the model
    classifies the images.}
\label{Fig: pred_data_ANN}
\end{figure*}

As a final step, we predicted the label of each image using the best
model found with the procedures previously described.  A set of 50 images
with their predicted labels is shown in
figure~(\ref{Fig: pred_data_ANN}).  Percentage values show the confidence
level with which the model can classify the images.  For the case of
this set of images, the model accurately predicted the labels of 42 images
and misread 8.  5 images of switching orbits were classified as circulation
cases, and 3 circulating orbits were labeled as switching ones.  All the
libration cases were correctly identified.  Values of the {\it Merit}
metric for libration, switching and circulation images were 1.000, 0.755,
and 0.877, respectively.

As a general rule in machine learning, the greater the size of the training
sample, the better the model performance. Classification of images with
{\it ANN} usually requires a training sample of the order of 60000 images
(see, for instance, the example of clothes images classification using the
Fashion MNIST data-set in the {\it Keras} documentation pages
\citet{Chollet_2018}).
For the case of the M1:2 asteroidal population, this is simply not viable,
since there are just 5700 numbered asteroids in the range of $a$
near the resonance ($2.411 < a < 2.426$ au), i.e., an order of magnitude
less.  Yet, despite this fundamental limitation, our model performs quite
well.  To quantitatively estimate its efficiency, we computed values
of our metrics, {\it Completeness, Purity} and {\it Merit}, for the same
set of 50 images of M1:2 resonant arguments, increasing the size
of the training set.  Values of these metrics were computed for
the three different types of orbits, libration, switching and circulation.
For all the simulations, {\it Accuracy} values were all above 0.996.

\begin{figure*}
  \centering
  \centering \includegraphics[width=5.5in]{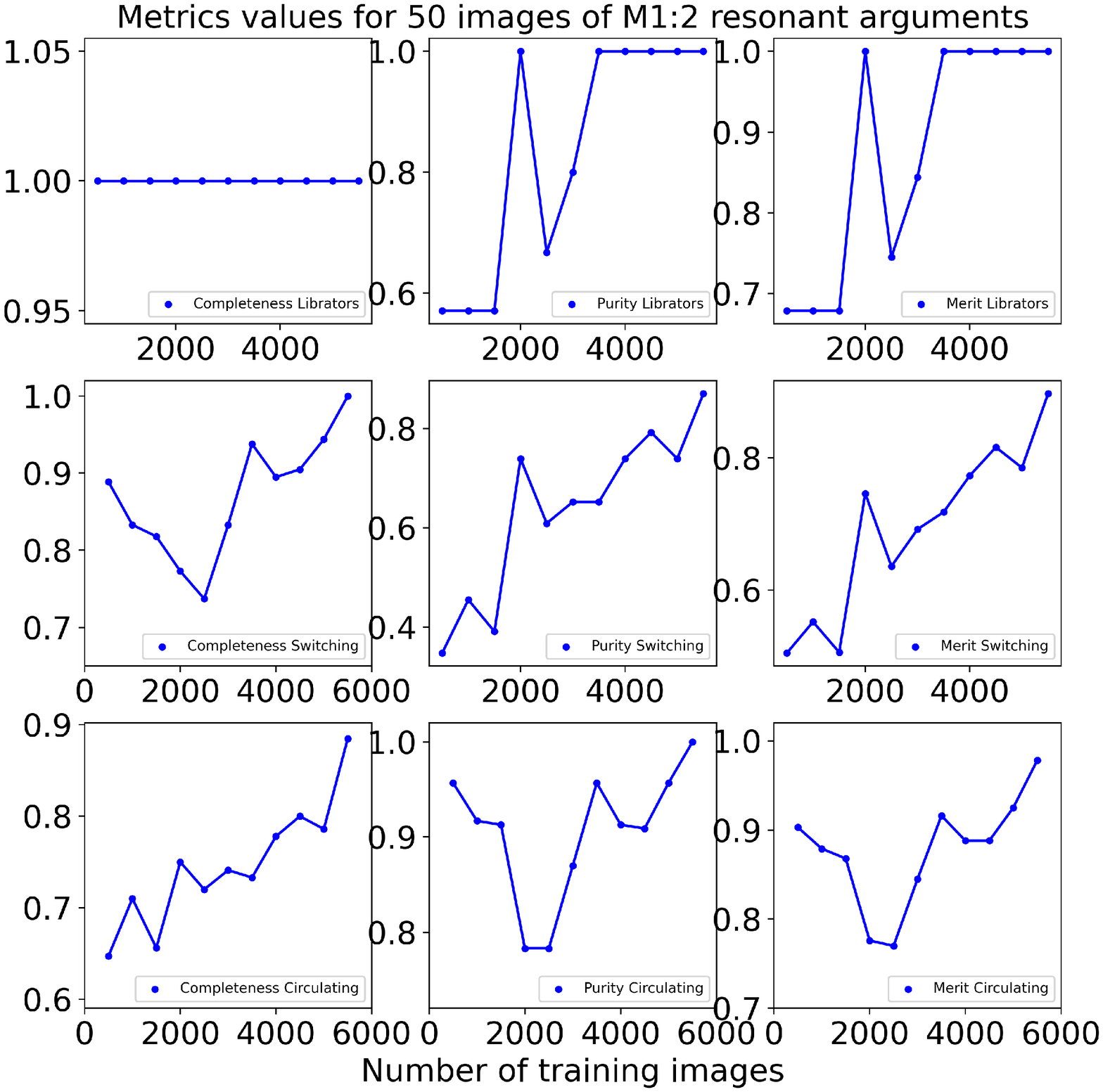}
  \caption{Values of {\it Completeness, Purity} and {\it Merit} for a set
    of 50 images of M1:2 resonant arguments for asteroids on librating,
    switching, and circulating orbits. The labels in each of the nine panels
    identify the metric value for each figure.}
\label{Fig: images_metrics_ANN}
\end{figure*}

Figure~(\ref{Fig: images_metrics_ANN}) displays our results.  Since the
results of {\it ANN} are inherently stochastic, individual data point can
change if we repeat the numerical experiment.  But the overall trends should be
robust.  The model {\it Merit} improves in all cases for increasing values of
the size of the training set.  The model can more easily identify images of
librating asteroids, since they are more distinguished from the other classes
of orbits.  Values of {\it Merit} reach 1.00 already for a training sample
of 3500 images.  The lowest performance was obtained for switching orbits, which
are easier to be confused with the other two classes. However, even for
this kind of orbits, the model could achieve values of {\it Merit} larger
than 0.80, if a sample large enough ($> 4500 $ images) is used.
Overall, {\it ANN} can be used to provide a preliminary classification
of asteroids resonant arguments, with good results.

\section{Applications of genetic algorithms to M1:2 resonant arguments labels}
\label{sec: gen_algo}

The next step of our analysis would be to predict the labels of asteroids
near the M1:2 resonance based on their proper elements distribution and
the labels of an appropriate training sample.  For this purpose, we
can either use a machine learning algorithm or an {\it ANN}.  As discussed
in the previous section, {\it ANN} become competitive with standard machine
learning approaches for large sizes of the training sample, which is not
the case for our problem. A possible application of {\it ANN}, and its
limitations, will be discussed in section~(\ref{sec: ANN_labels}).  Here,
we will focus our attention on standard machine learning approaches.

Machine learning methods, either if {\it standalone}, where a single
algorithm is applied, or {\it ensemble} methods, where several algorithms
are combined, depend on several model parameters, or hyper-parameters.
For instance, {\it Random Forest} methods that use several single
{\it Decision Trees} depend on the number of trees used, which is a
hyper-parameter that needs to be optimized.  Identifying the optimal
machine learning method and the combination of hyper-parameters for a
given problem may be a long and time-consuming process.  Here, as done
in \citet{Carruba_2021} for the case of asteroids near the $z_1$ and
$z_2$ secular resonances, we use an approach based on {\it genetic
  algorithms} \citep{Peng-Wei_2004}.

{\it Genetic algorithms} use an approach based on genetic evolution.
First, several models and their related combinations of hyper-parameters
are created.  After an iteration of the model, also called
{\it generation}, a scoring function can be used to identify the best
models.  Models similar to the best ones can then be created, and the
process can be repeated until some conditions are satisfied.
Interested readers can found more details on this procedure in
\citep{Peng-Wei_2004} and \citep{Carruba_2021}.  As in the last
paper, we used the {\it Tpot Python} library \citep{Trang_2020, Olson_2016}
with 5 {\it generations}, a {\it population size} (the
number of models to keep after each {\it generation}) of 20, and
a cross-validation {\it cv} equal to 5.  We also used thee values of
the random state: 42, 99, and 122, which correspond
to three different models: {\it XGBoost, GBoost,} and {\it Random Forest}.  The
specifications of the best among these models will be discussed later
on in this section.

\begin{figure}
  \centering
  \centering \includegraphics[width=3.5in]{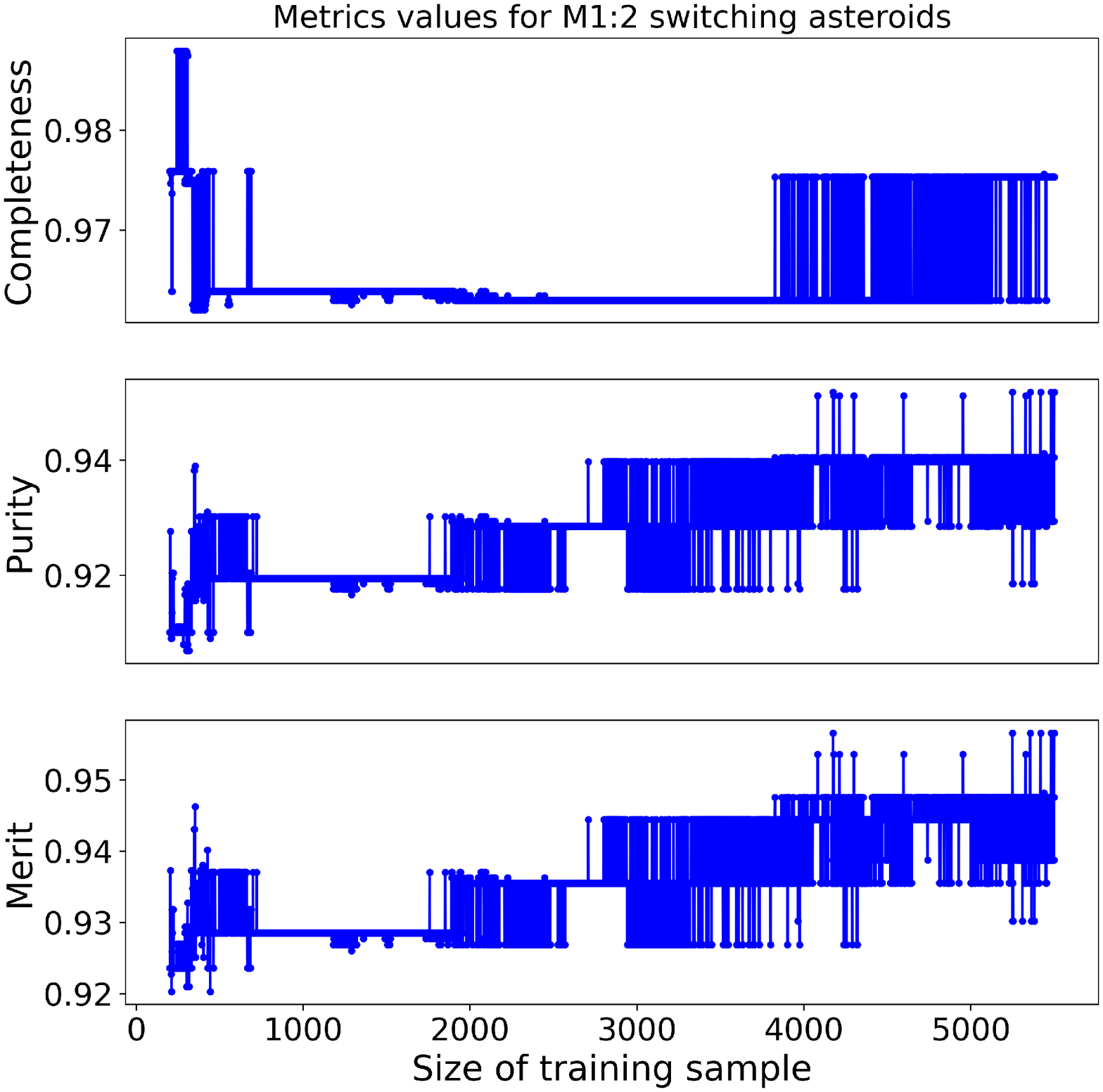}
  \caption{Values of {\it Completeness, Purity} and {\it Merit} for 
    test labelled asteroids on switching orbits as a function of the
    training sample size, obtained with the {\it Random Forest} algorithm.}
\label{Fig: Gen_metrics}
\end{figure}

To test these models we divided our sample of 5700 labelled asteroids into
three parts: a training set of 200 asteroids, a test set of 200 bodies,
and a pooling sample with the rest of the labelled objects.  The size
of the test sample is large enough for the results to be statistically
significant (3.51\% of the available data), but small enough to leave
space for enough data in the initial training and pooling sample.
A random asteroid
is selected in the pooling sample, added to the training set, and the model
is fitted to the test sample.  Values of the metrics are computed, and
the procedure is then repeated until there are no more objects in the
pooling sample.  Figure~(\ref{Fig: Gen_metrics}) displays the values of
{\it Completeness, Purity}, and {\it Merit} for the switching orbits class,
the type of orbits that previous analysis showed to be the most difficult
to predict, obtained by the best model among the tested ones, the
{\it Random Forest} algorithm.  This model and its hyper-parameters
are discussed in appendix 1.

The {\it Random Forest} reaches a plateau
in values of {\it Completeness, Purity}, and {\it Merit} for a training
size of $\simeq 3000$.  We will use this model to predict the labels of
unlabelled asteroids in section~(\ref{sec: res_groups})

\subsection{Applications of {\it ANN}}
\label{sec: ANN_labels}

{\it ANN} can also be applied to predict the labels of near resonance
asteroids.  However, as previously discussed the training sample
available for this problem is too small for this method to be used
advantageously.  We created a three-layered {\it Keras} model with an inner
layer of 200 neurons, one each for the asteroids in the test sample,
a hidden layer of 100 neurons, i.e. 50\% of the number of neurons in
the inner layer, as it is usually recommended, and an outer layer
of 3 neurons, one for each orbital-class. We choose to work with a training
sample of 5500 asteroids and a test sample of 200, to have a large
training sample, with 96.5\% of the available data. Other choices for
size of the test sample are, of course, possible.  We expect, however,
that the model results should be inferior for smaller sizes of the training
sample. We run this model over
100 epochs with a {\it callback} instruction, to identify the labels of the
same test sample used in section~(\ref{sec: gen_algo}).
The model was not able to identify librating asteroids, and values
of {\it Completeness, Purity} and {\it Merit} for circulating and switching
orbits were consistently below what predicted using {\it genetic algorithms}.
Given these considerations, we will not use {\it ANN} to predict
labels of unlabelled asteroids hereafter.

\section{Identification of resonant groups}
\label{sec: res_groups}

\begin{figure*}
\centering
\includegraphics[width=3.0in]{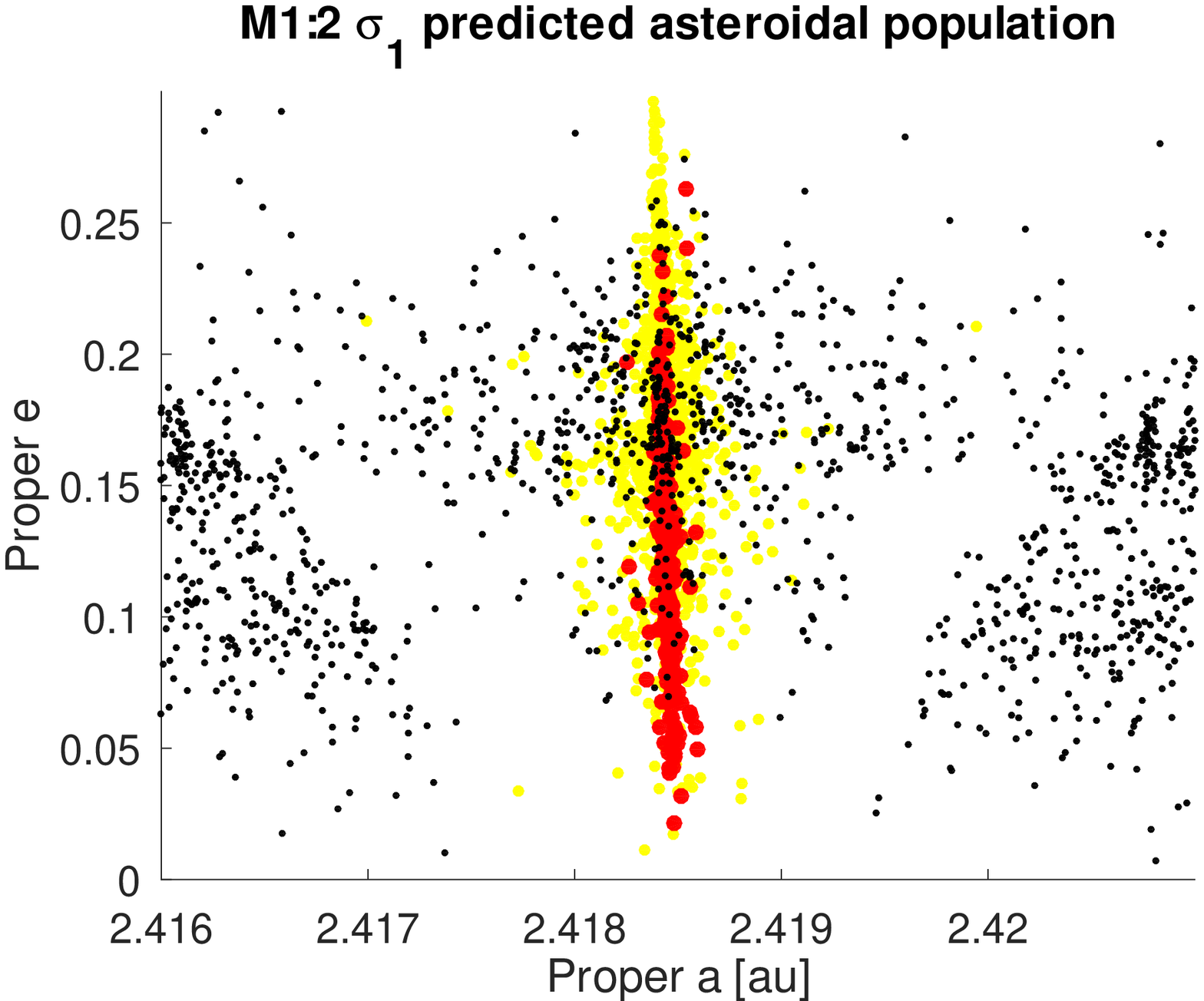}
\caption{A proper $(a,e)$ projection of asteroids in the region of
  the M1:2 resonance.  The colour code of predicted and confirmed
  asteroids in the region is the same as that of
  figure~(\ref{Fig: m12_expl_anal}).}
\label{Fig: M12_predicted}
\end{figure*}

Having identified the best performing supervised learning algorithm in
section~(\ref{sec: gen_algo}), here we use this method to predict the labels
of 740 multi-opposition asteroids, obtained from the $AFP$, using the
5700 asteroids that we previously classified as a training set.
Figure~(\ref{Fig: M12_predicted}) displays a proper $(a,e)$ projection
of 6440 asteroids for which we obtained labels, using the same colour
code as in figure~(\ref{Fig: m12_expl_anal}).  The predicted labels
are very consistent with those obtained in the preliminary analysis,
which confirms the validity of our method.

As a final check, we searched for possible dynamical clusters in the
populations of M1:2 asteroids on librating and switching orbits, to see
if our results are consistent with those in the literature.
\citet{Gallardo_2011} found that the three asteroids families most
affected by the M1:2 mean-motion resonance were those of Nysa, Massalia,
and Vesta.  Here, following the approach of \citet{Carruba_2021}, we
use learning Hierarchical Clustering Method (HCM), as implemented in
\citet{Carruba_2019}, on a domain of proper elements for the group
of M1:2 asteroids above described.  The procedure used to implement
this method was the same as that applied in \citet{Carruba_2021}:
a critical distance cutoff $\frac{1}{2} d_0$ was obtained, and
groups were identified for values of $d_0 \pm 5$~m/s.
We then verified if members of groups identified in this domain were
listed as members of a family by \citet{Milani_2014} and \citet{Nesvorny_2015}.
Please note that \citet{Milani_2014} reports the Nysa family as Hertha.
Our results are summarized in table~(\ref{Table: m12_groups}).
We selected groups that have at least 10 members at the critical distance
cutoff value, 5 members at the lowest distance cutoff of
$d_0 - 5 = 22.75$~m/s, and were still identifiable at the highest distance
cutoff of $d_0 - 5 = 32.75$~m/s.  Interested readers could find more
details on the procedures used in \citet{Carruba_2019, Carruba_2021}.
   
   \begin{table*}
      \begin{center}
        \caption{The table reports the dynamical groups with at least 10 members among the librating and switching M1:2 population, listed from the most to the least numerous, identified with the hierarchical clustering algorithm, at three values of the distance cutoff: (1) 22.75, (2) 27.75, and (3) 32.75 m/s.  The fourth column reports how many of the asteroids belong to a known asteroid family.}
        \label{Table: m12_groups}
         \begin{tabular}{|c|c|c|c|c|c|}
\hline
Family  & Number of  &  Number of  & Number of   & Family members      \\
Id.     & members (1) &  members (2) & members (3)  & with known fam. ID. \\   
\hline
19205 (1992 PT)    &   8  & 50   &  84  & Massalia: 28, Nysa:3    \\
42462 (5278 T-3)   &  13  & 19   &  25  & Massalia: 13, Nysa:1    \\
120135 (2003 GF7)  &  10  & 17   &  28  &     Nysa: 2             \\
95459 (2002 CF307) &   8  & 14   &  18  &    Vesta: 2             \\
44931 (1999 VD39)  &   7  & 10   &  10  &     Nysa: 4             \\
73066 (2002 FV15)  &   8  & 10   &  40  & Massalia: 8             \\
10516 Sakurajima   &   6  & 10   &  15  &     Nysa: 5, Massalia:1 \\
\hline
\end{tabular}
\end{center}
\end{table*}

Our analysis produced seven possible groups, all associated with the
Massalia, Nysa, and Vesta families, so confirming the analysis of   
\citet{Gallardo_2011}.
 
\section{Conclusions}
\label{sec: concl}

The main result of this work is the use of {\it ANN} for identifying
the behaviour of M1:2 resonant arguments images.  It is the first time,
to our knowledge, that {\it ANN}s have been used for such purpose in the
field of asteroid dynamics.  The use of this model allowed us to
classify the orbital type of all numbered asteroids in the orbital
region affected by this resonance, which has also been independently
confirmed by a visual analysis by all authors.

The labels for the population of numbered asteroids near the M1:2 mean-motion
resonances were also used to predict the orbital status of multi-opposition
asteroids.  Using genetic algorithms, we identify the best performing
supervised learning method for our data, that we used to obtain labels
for asteroids, without the need to perform a numerical simulation and
an analysis of resonant angles.

The identification of clusters in the population of asteroids in librating
and switching orbits suggested that three asteroid families, those of
Massalia, Nysa, and Vesta, are the most dynamically affected by this resonance,
so confirming the analysis of previous authors \citep{Gallardo_2011}.

The methods developed in this work could be easily used for other cases
of asteroids affected by mean-motion resonance, like the ones studied by 
\citet{2017MNRAS.469.2024S}.  We consider these models as the main result
of this work.

\section{Appendix 1: Genetic algorithms outcome}
\label{sec: Appendix 1}

The best performing model provided by {\it genetic algorithm} was the
{\it Random Forest} algorithm.  As described in \citep{Carruba_2021},
this algorithm is an ensemble method that uses several standalone
{\it decision trees}.  The training data can be divided into multiple
samples, the bootstrap samples, that can be used to train an
independent classifier.  The outcome of the method is based on a majority
vote of each {\it decision tree}. Important parameters of
this model,  as described in \citet{Swamynathan_2017}, are:

\begin{enumerate}
\item {\it Bootstrap}: Whether the algorithm is using bootstrap samples ({\it True}) or not ({\it False}).
\item {\it Criterion}: The function to measure the quality of a split. The supported criteria are “{\it gini}” for the Gini impurity and “{\it entropy}” for the information gain.
\item {\it max$\_$features}: The random subset of features to use for each splitting node.
\item {\it min$\_$samples$\_$leaf}: The minimum number of samples required to be at a leaf node. 
\item {\it min$\_$samples$\_$split}: The minimum number of data points placed in a node before the node is split.
\item {\it Number of estimators}: the number of decision trees algorithms.    
\end{enumerate}
   
Our model used {\it Bootstrap = True}, a {\it gini Criterion}, 
{\it max$\_$features = 1.0}, {\it min$\_$samples$\_$leaf = 12},
{\it min$\_$samples$\_$split = 18}, and 100 {\it estimators}.

\section*{Acknowledgments}

We are very grateful tp the reviewer of this paper, Dr. Evegny Smirnov,
for helpful and constructive comments that improved the quality of this paper.
We would also like to thank the Brazilian National Research Council
(CNPq, grant 301577/2017-0) and The Coordination for the Improvement of
Higher Education Personnel (CAPES, grant 88887.374148/2019-00).
We acknowledge the use of data from the Asteroid Dynamics Site ($AstDys$,
\citet{Knezevic_2003}, http://hamilton.dm.unipi.it/astdys) and the
Asteroid Families Portal ($AFP$, \citet{Radovic_2017},
http://asteroids.matf.bg.ac.rs/fam/properelements.php). Both databases
were accessed on August 1st 2020.
VC and WB are part of "Grupo de Din\^{a}mica Orbital \& Planetologia (GDOP)"
(Research Group in Orbital Dynamics and Planetology) at UNESP, campus
of Guaratinguet\'{a}.
This is a publication from the MASB (Machine-learning applied to small bodies, \\ https://valeriocarruba.github.io/Site-MASB/) research group.  Questions on
this paper can also be sent to the group email address: {\it mlasb2021@gmail.com}.

 \section{Availability of data and material}

 All image data on numbered asteroids near the  M1:2 resonance is available
 at:\\
 https://drive.google.com/file/d/1RsDoMh8iMwZhD-fnkYSs9hiWmg96SZf0/view?usp=sharing. 

 \section{Code availability}

 The code used for the numerical simulations are part of the {\it SWIFT}
 package, and are publicly available at: \\*
 https://www.boulder.swri.edu/~hal/swift.html, \citep{levison_1994}.\\
 Deep learning codes were written in the {\it Python} programming
 language and are available at the {\it GitHub} software repository,
   at this link:\\

 https://github.com/valeriocarruba/ANN\_Classification\_of\-\_M12\_resonant\_argument\_images\\

 Any other code described in this paper can be obtained from the first author
 upon reasonable request.
 
\bibliographystyle{mnras}
\bibliography{mybib}

\bsp
\label{lastpage}

\end{document}